\documentclass[article, prX]{revtex4}
\usepackage{graphicx}
\begin{document}
\title{Tracking the dynamics of skyglow with differential photometry using a digital camera with fisheye lens}

\author{Andreas Jechow$^{1,2}$, Salvador J. Ribas$^3$, Ramon Canal-Domingo$^3$, Franz H{\"o}lker$^1$, Zolt{\'a}n Koll{\'a}th$^4$, Christopher C. M. Kyba$^{2,1}$}
\affiliation{$^1$ Ecohydrology, Leibniz Institute of Freshwater Ecology and Inland Fisheries, M{\"u}ggelseedamm 310, 12587 Berlin}
\affiliation{$^2$ Remote Sensing, Helmholtz Center Potsdam, German Center for Geosciences GFZ, Telegraphenberg, Potsdam}
\affiliation{$^3$ Parc Astron{\`o}mic Montsec, Comarcal de la Noguera, Pg. Angel Guimer{\`a} 28-30, 25600 Balaguer, Lleida, Spain}
\affiliation{$^4$ E{\"o}tv{\"o}s Loránd University, Savaria Department of Physics, K{\'a}rolyi G{\'a}sp{\'a}r t{\'e}r 4, 9700 Szombathely, Hungary}

\maketitle
\section*{Abstract}
Artificial skyglow is dynamic due to changing atmospheric conditions and the switching on and off of artificial lights throughout the night. Street lights as well as the ornamental illumination of historical sites and buildings are sometimes switched off at a certain time to save energy. Ornamental lights in particular are often directed upwards, and can therefore have a major contribution towards brightening of the night sky. Here we use differential photometry to investigate the change in night sky brightness and illuminance during an automated regular switch-off of ornamental light in the town of Balaguer and an organized switch-off of all public lights in the village of \`{A}ger, both near Montsec Astronomical Park in Spain. The sites were observed during two nights with clear and cloudy conditions using a DSLR camera and a fisheye lens. A time series of images makes it possible to track changes in lighting conditions and sky brightness simultaneously. During the clear night, the ornamental lights in Balaguer contribute over 20\% of the skyglow at zenith at the observational site. Furthermore, we are able to track very small changes in the ground illuminance on a cloudy night near \`{A}ger.

\section*{1. Introduction}
Light in nature is very dynamic, with ground illuminance varying over nine orders of magnitude from more than 100,000 lx during a clear day to about 100 $\mu$lx during an overcast night \cite{seidelmann2005explanatory}. Until the introduction of artificial lighting, the light present in the environment was due to celestial light sources and the atmosphere. The regular diurnal and seasonal patterns in light and darkness caused by the motion of the Earth and Moon shaped biological rhythms during evolution \cite{vosshall1994block}, and changes in light are used as clock synchronizers for animals to anticipate times for reproduction, rest or foraging \cite{elangovan2001effect,kronfeld2013chronobiology}.

With the invention of artificial lights, humans can decorrelate their activities from these natural patterns and extend productive working hours. Artificial light at night (ALAN) comes with the downside of light emission into the primordially dark environment. The resulting light pollution \cite{Riegel1973} can have diverse effects on flora, fauna and humans \cite{Longcore2004,book:rich_longcore,Stevens:2015,Gaston:2015_ptb,gaston2013ecological,Hoelker:2010_b}. A recent study with satellite data showed that ALAN is growing in area and intensity by more than 2 $\%$ per year globally, with individual countries having increases of more than 10 $\%$ per year \cite{kyba2017VIIRS}. If ALAN is directly or indirectly radiated towards the sky it can be scattered or reflected within the atmosphere, leading to an increased night sky brightness (NSB) \cite{Aube:2015}. This artificial skyglow is the form of light pollution with an impact on a large spatial scale \cite{falchi2016WA,Kyba:2015_isqm,Hoelker:2010_a}. The intensity and spatial extent of skyglow change with atmospheric conditions, most dramatically with clouds \cite{Kyba:2015_isqm,Kyba:2011_sqm,Ribas2016clouds,jechow2017balaguer}.

ALAN itself can be highly dynamic as lights are switched on and off throughout the night. Dobler et al. used these dynamics to study aggregate human behavior, by analyzing a time series of images of the New York skyline \cite{dobler2015dynamics}. ALAN dynamics are also coupled to the NSB as strikingly shown in a recent study by Bara et al. \cite{bara2017estimating}, enabling them to disentangle the contribution of specific ALAN sources to skyglow.

Over the last decades, it has become increasingly popular to illuminate historical buildings for aesthetic purpose. Unfortunately, these ornamental lights are frequently poorly designed, with lights shining upwards, often missing the target by a large fraction, and radiating at overly high light levels. This light spill into the sky supposedly contributes largely to skyglow, that has several other (indirect) components like light reflected off from the ground or buildings.

In this work, we study the dynamics of skyglow with a fisheye lens and a digital single lens reflex (DSLR) camera. We observe two switch-offs, each during both a clear and a partly cloudy night. The regularly scheduled switch-off of ornamental light in the town of Balaguer was observed with the camera obtaining all-sky images in the horizontal plane. The switch-off of all public light in the village of \`{A}ger was organized for an intercomparison campaign of light measurement devices \cite{RibasLonne2017}, and was observed with imaging in the vertical plane from a vantage point (lookout). Among the many methods to measure the NSB (see \cite{hanel2017measuring} for a recent overview), we chose DSLR cameras because they are straightforward to operate and highly mobile \cite{Kollath:2010,kollath2017night,Jechow2016,jechow2017measuring,jechow2017balaguer}. Our results show further potential and limitations of wide-angle differential photometry, which can be used to remotely identify sources of light pollution.

\section*{2. Methods}
\subsection*{2.1. Study site and switch-offs}
The data was obtained during the 2016 Stars4All/LoNNe Intercomparison Campaign held at Montsec Astronomical Park (Parc Astron\`{o}mic Montsec, Centre d'  Observaci\'{o} de l'Univers: PAM-COU). Measurements were performed during a clear night (May 3rd to May 4th) and a partly cloudy night (May 5th to May 6th). A set of data acquired on the same nights was published in another paper dealing with the spatial extent of skyglow and its change with clouds \cite{jechow2017balaguer}. Full details about the Intercomparison Campaign can be found in the campaign report \cite{RibasLonne2017}.

The Municipality of Balaguer provided a brief summary of installed lamps and the time that ornamental lights were turned off. The village of \`{A}ger allowed us to turn all street lighting off at a time of our choice. The automatic switch-off of ornamental lights was observed from the outskirts of Balaguer. Then a transect was conducted along the highway C-12 leading North towards \`{A}ger, as reported in \cite{jechow2017balaguer}. The last stop of the transect was at Port d' \`{A}ger, where the organized switch-off was observed. The topography of the area is dominated by the mountains of the Montsec region. According to the global skyglow model from Falchi et al. \cite{falchi2016WA}, the clear sky luminance at the zenith in the center of Balaguer is 1.59 mcd/m$^2$ and 0.25 mcd/m$^2$ in \`{A}ger. See Appendix A for annual NSB plots of Balaguer and PAM-COU close to \`{A}ger both measured at zenith with Sky Quality Meters (SQMs), commonly used photometers (Unihedron, Canada) \cite{Ribas2016clouds,Kyba:2015_isqm,hanel2017measuring}.
\begin{figure}[tp]
\centering
\includegraphics[width=0.8\columnwidth]{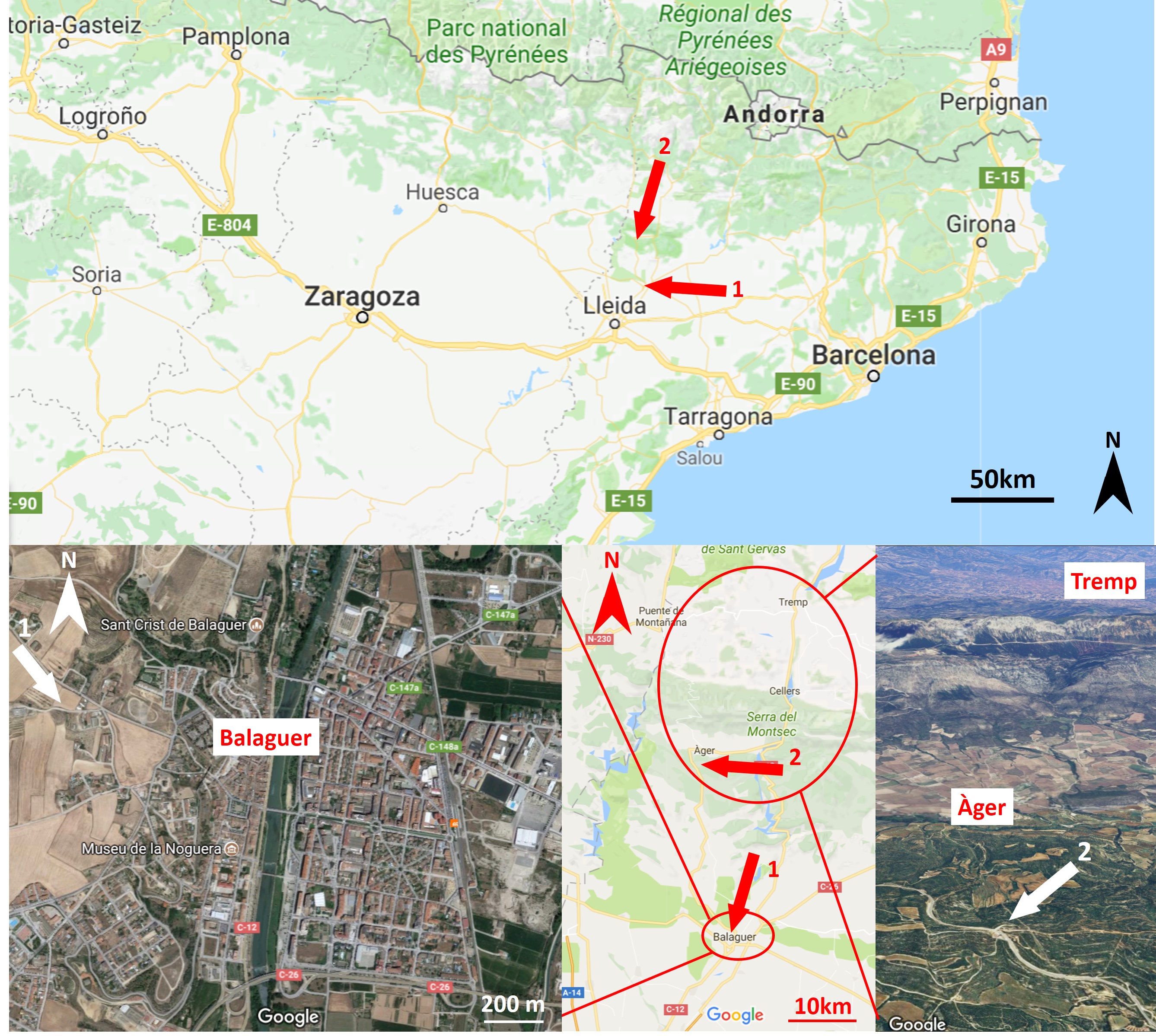}
\caption{Maps of the observation locations, upper map shows parts of Spain, lower left hand side shows observation site in Balaguer (1) and lower right hand side in Port d' \`{A}ger (2). See text for details. (Maps from Google Maps)}
\label{map}
\end{figure}

\subsubsection*{Ornamental lights in Balaguer}
The town of Balaguer (ca. 16,500 inhabitants) is located in Catalonia, Spain, about 25 km northeast of the city of Lleida (ca. 138,000 inhabitants) and about 120 km northwest of Barcelona. The town is nestled on the southern edge of the Montsec mountain region, about 27 km south of Montsec Astronomical Park. The observation point (41.7911N, 0.7975E, 273m elevation) was located about 1 km west of the city centre (see Fig. \ref{map} left hand side with label 1), at a position with no nearby lamps that would have affected the measurement. The new World Atlas of artificial NSB \cite{falchi2016WA} reports a clear sky luminance at zenith of 1.16 mcd/m$^2$ at this site \cite{falchi2016supplement}.

According to a document (recent at the time of our measurement campaign) provided to us by the Municipality of Balaguer, 3131 lamps were installed in the city, with a total power of 616.4 kW and an annual operation time of about 4226 hours. 2281 of these lamps were high-pressure sodium, 400 were metal halide lamps, 252 LEDs ($<$3500 K CCT) and the rest were other technologies. Most of the public lights were cut-off (ca. 2180), ca. 215 were semi cut-off and the remaining 733 were non cut-off or ornamental. The observed switch-off comprised 280 lamps, which is about 9$\%$ of the 3131 lamps. The automated switch-off was scheduled at 23:30 local time (21:30 UTC) for both nights, with an extra manual switch-off performed between 23:30 and 23:42 local time during the first night.

\subsubsection*{Public lights in \`{A}ger}
The village of \`{A}ger (42.0024N, 0.7608E, 607m elevation, ca. 415 inhabitants in core village) is located about 23 km north of Balaguer and 3.5 km southeast of the PAM-COU of Montsec Astronomical Park. Fortunately for astronomical observations, the PAM-COU is located in a valley surrounded by hills and mountains, which tends to shield the artificial skyglow present at the horizon. For the majority of clear nights, the zenith night sky brightness is near-natural with values of 22 mag$_{SQM}$/arcsec$^2$ \cite{Ribas2016clouds}. On cloudy nights the zenith night sky brightness can either decrease or increase, as shown by an in depth-study linking SQM values and cloud height at PAM-COU \cite{Ribas2016clouds}.

In \`{A}ger, 185 high pressure sodium lamps with ULOR (upper light output ratio) $<$1$\%$ are installed. The power of the lamps ranges from 70 to 150 W, most commonly between 70 to 100W. In addition to this public lighting, there are about 12 privately owned lamps, that remained switched on. 

During the first night of observation, the town of \`{A}ger switched off lights sequentially starting at 02:40 local time with the last section being switched off at 02:52 local time on May 4th. During the second night of observation, the lights were turned off between 02:06 local time and 02:14 local time on May 6th.
\begin{figure}[tp]
\centering
\includegraphics[width=0.55\columnwidth]{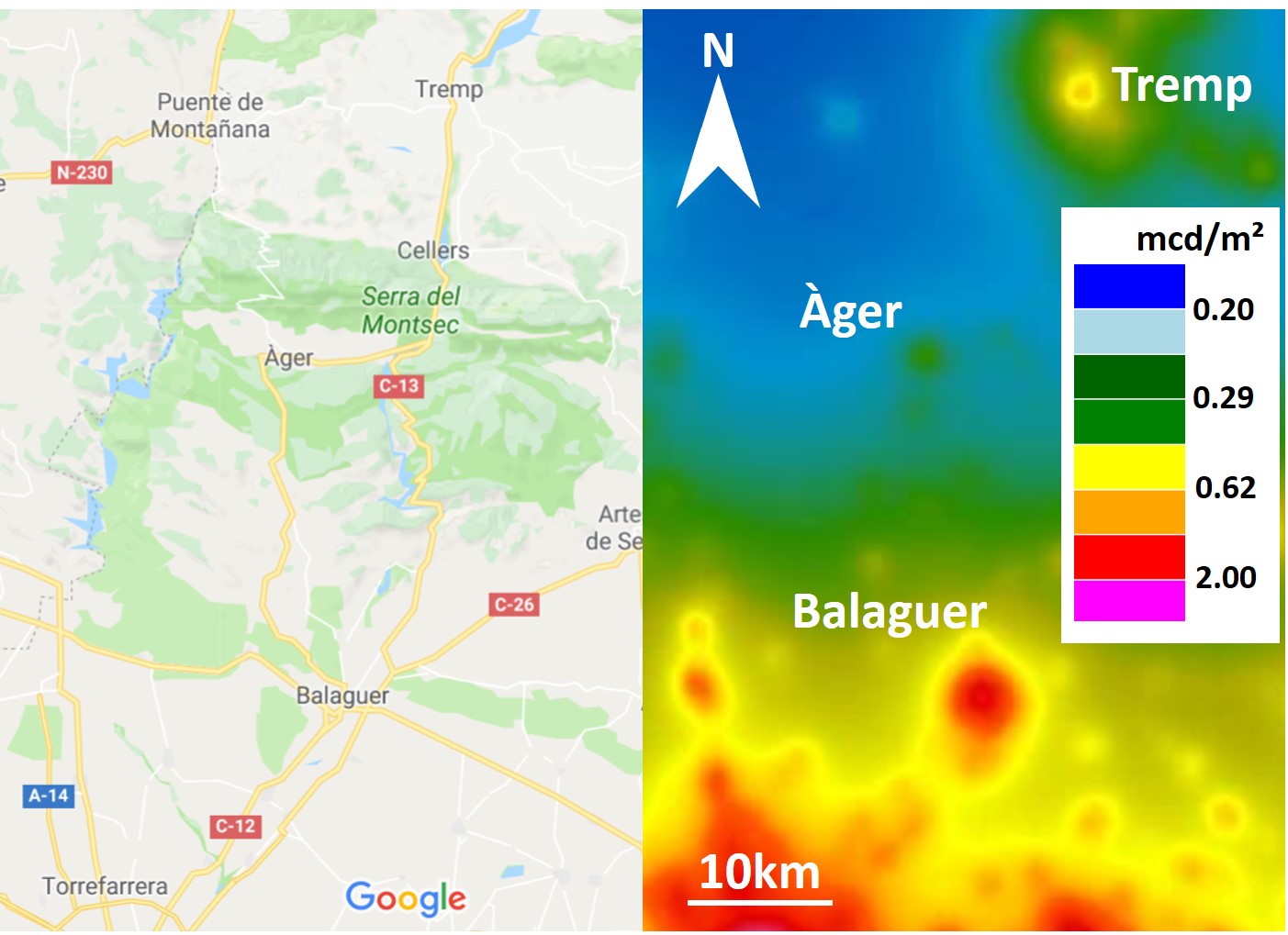}
\caption{Maps of the observation locations and surroundings showing the simulated NSB of the area. Data on right hand side is taken from \cite{falchi2016WA,falchi2016supplement}, right hand side map by Jurij Stare, www.lightpollutionmap.info. (Left hand side map from Google Maps)}
\label{WA}
\end{figure}

The switch-offs were observed at a public lookout Port d' \`{A}ger (41.9791N, 0.7506E, 908m elevation) in Port d' \`{A}ger, where it is possible to overlook the valley of \`{A}ger (see Fig. \ref{map} right hand side with label 2). The new World Atlas of artificial NSB \cite{falchi2016WA} reports a clear sky luminance at zenith of 0.22 mcd/m$^2$ at this site \cite{falchi2016supplement}. The vantage point is located at an elevation 300 m higher than the village, at about 2.5 km distance from the village of \`{A}ger.

\subsection*{2.2. Weather conditions}
On May 3rd 2016, the sky was clear during the day with some cirrus, still significant around sunset but disappearing by around nautical twilight. No clouds were observed by the ceilometer (Jenoptik CHM15K, a LIDAR light detecting and ranging based ceilometer equipped with an IR-laser) at PAM-COU after 22:20 local time (20:20 UTC). At PAM-COU, the sky was cloud-free by astronomical twilight, and this condition lasted through the entire night, including during the morning twilight. In Balaguer, very few thin clouds were observed at the beginning of data taking.

On May 5th, the sky became progressively cloudier throughout the day, with high clouds. The entire evening was characterized by high thin clouds. At PAM-COU, stars were visible, and there was a moisture layer up to the top of the mountain which allowed scattered light from very remote cities to be seen. The sky was covered the whole night with medium and high clouds. The ceilometer measured cloud heights of about 6 km for the relevant measurement period. Early in the morning low clouds also appeared, according to ceilometer measurements.

\subsection*{2.3. DSLR camera and software}
Commercial DSLR cameras with fisheye lenses provide a simple and robust method for NSB monitoring. These devices are less rigid than commercial outdoor and weatherproof light measuring devices like luxmeters, but still applicable in the field \cite{jechow2017balaguer}, even under the challenging condition of observations from a boat \cite{jechow2017measuring}. The advantages of DSLR cameras are the easy use, quick operation, moderate price and spatially resolved data in three spectral bands \cite{hanel2017measuring}.

We obtained fisheye images with a commercial DSLR camera, a Canon EOS 6D, which has a full-frame CMOS sensor with 20.2 Megapixel (5496 x 3670 pixels). The camera has a built-in GPS sensor, and allows ISO settings from 100 to 25,600 and shutter speed ranges down to 30 s in automatic mode. We operated the camera with a circular fisheye lens (Sigma EX DG with 8 mm focal length) always at full aperture of 3.5. On clear nights, we use a bright celestial object in the live view of the camera to focus the lens. In cloudy or overcast conditions, focusing is done using an artificial light source at large distance (e.g. on the horizon). For the time series images, the camera was mounted on a tripod. To acquire all-sky images, the camera was aligned with the center of the lens oriented towards the zenith, for vertical images the camera was aligned approximately pointing towards the horizon. In Balaguer, ISO 1600 was used and the shutter speed was set to 30 s for the clear night and 15 s for the cloudy night. In \`{A}ger, ISO 3200 and shutter speed of 30 s was used.

For image processing we used the commercial "Sky Quality Camera" software (Version 1.8, Euromix, Ljubljana, Slovenia). The camera is calibrated by the software manufacturer. According to the manufacturer, the calibration includes photometric calibration using the green channel of the camera, as well as correction of optical aberrations like vignetting. A cross check with the free software DiCaLum \cite{kollath2017night} that we used in our previous analysis \cite{jechow2017balaguer}, showed only slight deviation below 10 percent with DiCaLum yielding higher values. An in depth intercomparison between the two software and calibration methods will be topic of a planned future paper. The software makes it possible to process luminance maps, subtract images, obtain illuminance data, analyze sectors of the image and to do cross sections. The software calculates the cosine corrected illuminance $E_{v,cos}$ in the imaging plane:
\begin{equation}
E_{v,cos}=\int_{0}^{\pi \over 2}\int_{0}^{2\pi} L_{v,sky}(\theta, \phi) sin\theta cos\theta d\phi d\theta,
\end{equation}
and the scalar illuminance for the imaging hemisphere $E_{v,scal,hem}$ without cosine correction \cite{duriscoe2016photometric}:
\begin{equation}
E_{v,scal,hem}=\int_{0}^{\pi \over 2}\int_{0}^{2\pi} L_{v,sky}(\theta, \phi) sin\theta d\phi d\theta.
\end{equation}
In the equations, $L_{v,sky}$ is the sky luminance, $\theta$ is the zenith angle and $\phi$ is the azimuth angle. For all-sky images, i.e. when imaging in the horizontal plane, $E_{v, cos}$ is usually termed horizontal illuminance and for imaging in the vertical plane, $E_{v, cos}$ is usually termed vertical illuminance. For each image of the time series, we extract zenith luminance, and cosine corrected and (hemispherical) scalar illuminance as well as luminance from regions of interest.

\section*{3. Results and discussion}
\subsection*{3.1. Ornamental lights in Balaguer (Automated switch-off)}
\subsubsection*{Luminance maps}
\begin{figure}[tp]
\centering
\includegraphics[width=0.75\columnwidth]{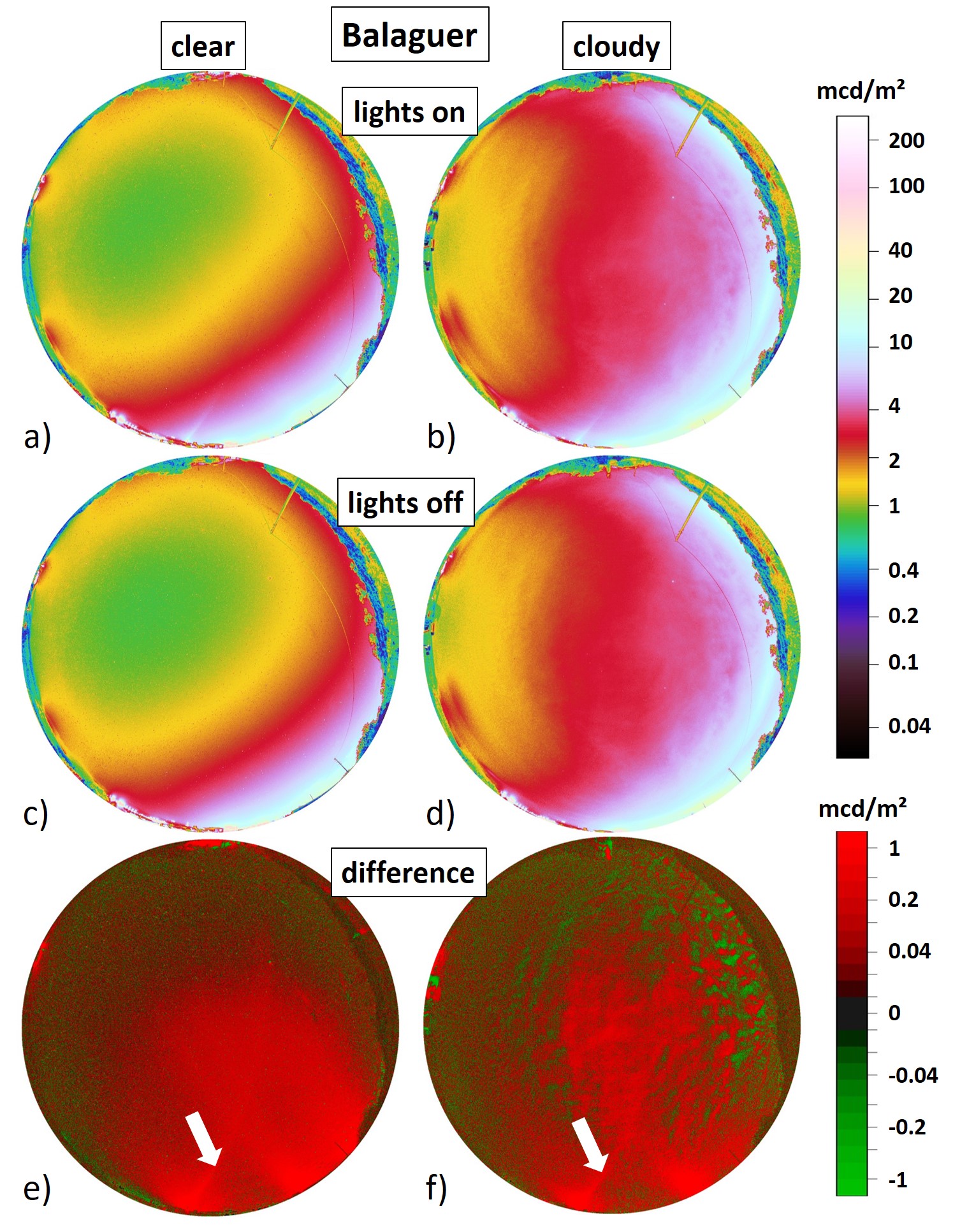}
\caption{Excerpt from the data obtained for the automated switch-off of ornamental lights in Balaguer, Spain. Left column (a, c, e) shows clear night data and right hand column (b, d, f) shows cloudy night data. The upper row (a, b) shows all-sky luminance maps before the switch-off. The middle row (c, d) shows all-sky luminance maps after the switch-off. The lower row shows differential luminance maps, obtained by subtracting the middle row data from the upper row data. White arrows indicate sky beamer position. See text for more details on times and location.}
\label{Bal_allsky}
\end{figure}

Figure \ref{Bal_allsky} shows part of the data obtained in Balaguer for the automated switch-off. The left column (a, c, e) shows data from the clear night and right column (b, d, f) data from the cloudy night. The upper row (a, b) shows luminance maps before the ornamental lights were switched off with images obtained on May 3rd 2016 starting at 23:29:51 and on May 5th 2016 at 23:29:44 local time, respectively. The middle row (c, d) shows luminance maps after the light has been switched off, obtained at May 3rd 2016, 23:32:06 and on May 5th 2016 at 23:30:22 local time, respectively. The lower row shows differential luminance maps, obtained by subtracting the images after the switch-off (middle rows) from the images before the switch-off (upper rows).

For both nights, the NSB distribution is heterogeneous with a bright region in the lower right corner of the image and a darker region in the upper left corner, respectively. The difference in the NSB between clear and cloudy conditions (left vs. right column) is immediately visible as discussed in detail in \cite{jechow2017balaguer}. When comparing luminance maps obtained on the same nights, the change in NSB due to switching of the lights is not as strikingly apparent as the change of the NSB due to clouds, but a closer look reveals differences. In both images obtained with the ornamental lights turned on (Fig. \ref{Bal_allsky} a, b), a small beam of light is apparent at the lower left side of the images, which is included in the automated switch-off and therefore not visible in the images with lights off (Fig. \ref{Bal_allsky} c, d). Furthermore, the bright region in the lower right corner is reduced in both images.
\begin{figure}[tp]
\centering
\includegraphics[width=0.35\columnwidth]{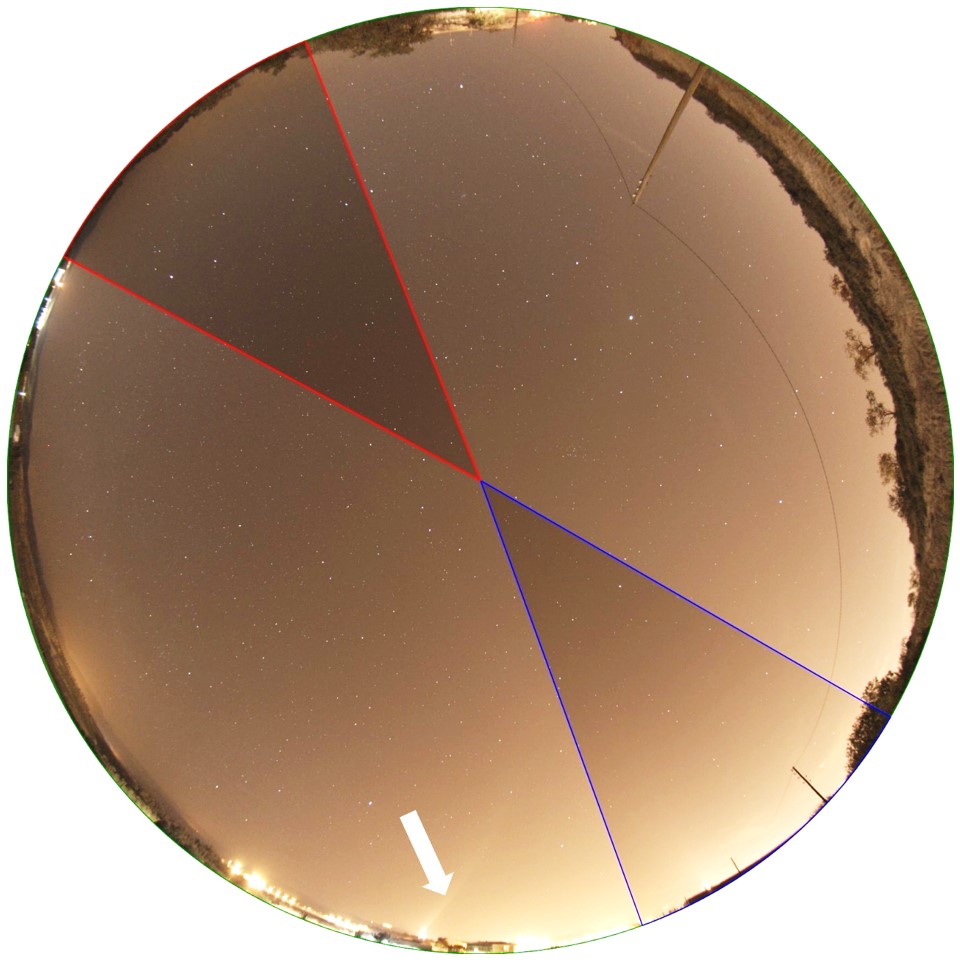}
\caption{RGB image obtained at the site in Balaguer on the clear night before the switch-off (Fig. \ref{Bal_allsky} a) showing the two regions with 40$^{\circ}$ width in azimuth angle where the luminance as function of zenith angle were analyzed. The red area represents the region between 20$^{\circ}$ and 60$^{\circ}$ azimuth angle pointing away from the town. The blue area represents the region region between 200$^{\circ}$ and 240$^{\circ}$ azimuth angle facing the town.}
\label{Bal_RGB}
\end{figure}

We used differential photometry to unravel small changes in the NSB, by subtracting the image after the switch-off from the image obtained before the switch-off. This is shown for the clear night in the lower left image, Fig \ref{Bal_allsky} e), and for the cloudy night in the lower right image, Fig. \ref{Bal_allsky} f). In these plots, red regions were brighter before the switch-off and green regions darker. In the differential luminance plots of both clear and cloudy night data, it is evident that most parts of the sky got darker with lights being switched off (red areas) and several light sources that shine light upwards can be identified by the red beams (indicated by white arrows). For the clear night data shown in Fig. \ref{Bal_allsky} e), few pixels got brighter after the switch-off (green color), which is mainly due to the rotation of the Earth and the changed position of the stars between the individual images. For the cloudy night data shown in Fig. \ref{Bal_allsky} f) more parts of the sky got brighter with switching off the lights (e.g. upper right corner). This is attributed to the moving clouds, as the sky was not completely overcast.
\begin{table}[bp]
  \centering
  \caption{Zenith luminance, horizontal and scalar illuminance for the two nights before and after the switch-off in Balaguer.}
  \label{table_zenith}
	\vspace{4mm}
  \begin{tabular}{c}
Zenith Luminance $ L_{v,zen}$ [mcd/m$^2$]
  \end{tabular}\\
  \begin{tabular}{ccccc}
 & lights on & lights off & $\Delta$& ratio\\
\hline
clear & 1.15 $\pm$ 0.12& 0.99 $\pm$ 0.10 & 0.16 & 1.16 \\
cloudy & 3.24 $\pm$ 0.3 & 2.96 $\pm$ 0.3 & 0.27 & 1.09 \\
\hline
\vspace{0.1cm}
  \end{tabular}\\
  \begin{tabular}{c}
Horizontal illuminance $ E_{v,hor}$ [mlux]
  \end{tabular}
	\linebreak
  \begin{tabular}{ccccc}
 & lights on & lights off & $\Delta$& ratio\\
\hline
clear & 5.36 $\pm$ 0.54& 4.92 $\pm$ 0.49 & 0.44 & 1.09 \\
cloudy & 10.86 $\pm$ 1.09 & 10.43 $\pm$ 1.04 & 0.43  & 1.05 \\
\hline
\vspace{0.1cm}
  \end{tabular}\\
    \begin{tabular}{c}
Scalar illuminance $ E_{v,scal,hem}$ [mlux]
  \end{tabular}\\
  \begin{tabular}{ccccc}
 & lights on & lights off & $\Delta$& ratio\\
\hline
clear & 14.65 $\pm$ 1.47& 13.16 $\pm$ 1.32 & 1.49 & 1.11 \\
cloudy & 24.17 $\pm$ 2.42 & 23.23 $\pm$ 2.32 & 0.94 & 1.04 \\
\hline
  \end{tabular}
\end{table}

\subsubsection*{Illuminance and angular luminance distribution}
We calculated plane illuminance, scalar illuminance, zenith luminance and the luminance as a function of the zenith angle for two regions indicated in Fig. \ref{Bal_RGB}. Table \ref{table_zenith} summarizes the values acquired before and after the switch-off for the zenith luminance (averaged over a 10$^{\circ}$ circle), horizontal and scalar illuminance. The zenith luminance for the clear night changed by 16$\%$ from 1.15 mcd/cm$^2$ to 0.9 mcd/cm$^2$. Please note that the value before the switch-off is in good agreement with the value of 1.16 mcd/cm$^2$ according the Falchi et al. \cite{falchi2016WA}. The difference in luminance was 0.16 mcd/cm$^2$, which is of similar value to the natural sky background brightness (which is about 0.17 mcd/cm$^2$, not including starlight) \cite{hanel2017measuring}. The contribution of the ornamental lights on the artificial skyglow can be estimated by assuming a near natural NSB (including starlight) of 0.25 mcd/cm$^2$ \cite{hanel2017measuring} and subtracting this value from the measured luminance values. Then the additional luminance due to skyglow is 0.90 mcd/cm$^2$ before and 0.74 mcd/cm$^2$ after the lights were switched off. The ornamental lights are therefore responsible for 21\% of the artificial component of the zenith NSB during the clear night at this location before the switch-off.

\begin{figure}[tp]
\centering
\includegraphics[width=0.8\columnwidth]{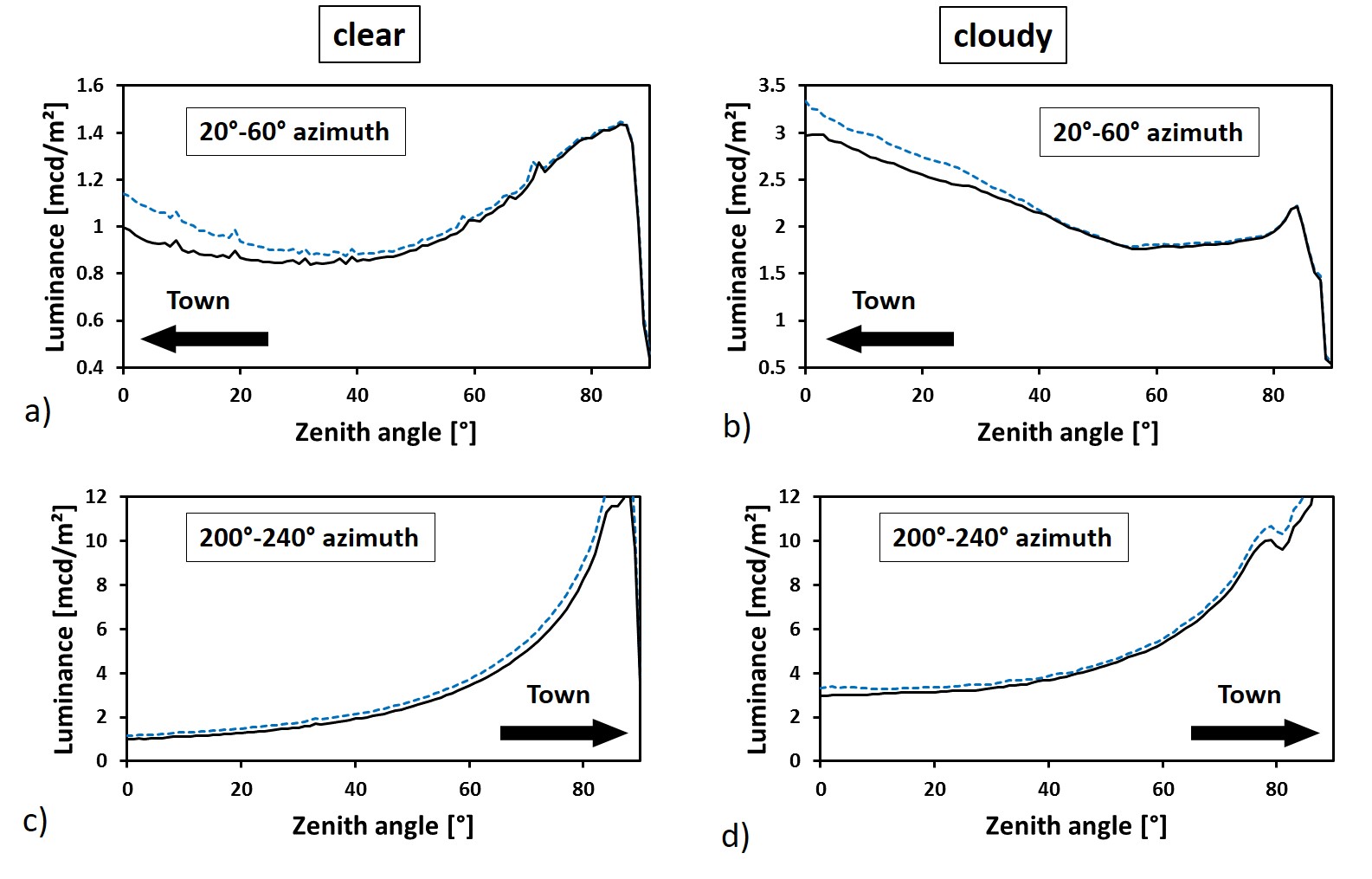}
\caption{Angular luminance distribution for clear night (left column) and cloudy night (right column) obtained from regions with 40$^{\circ}$ width in azimuth angle (see Fig. \ref{Bal_RGB}). The upper row shows a region between 20$^{\circ}$ and 60$^{\circ}$ azimuth angle pointing away from the town. The lower row shows a region between 200$^{\circ}$ and 240$^{\circ}$ azimuth angle facing the town. The blue dashed line in the plots show the luminance before and the black solid line after the lights were switched off. Black arrows indicate direction to town of Balaguer.}
\label{Bal_angle}
\end{figure}

The luminance is shown as a function of zenith angle in Fig. \ref{Bal_angle}, with the left column (a, c) showing again data from the clear and right column (b, d) showing data from the cloudy night. From the luminance maps, two regions with 40$^{\circ}$ width in azimuth have been processed, one ranging from 20$^{\circ}$ - 60$^{\circ}$ azimuth angle facing the town of Balaguer (red region in Fig. \ref{Bal_RGB}) and the other ranging from 200$^{\circ}$ to 240$^{\circ}$ azimuth angle pointing away from the town (blue region in Fig. \ref{Bal_RGB}). The upper row (a, b) in Fig. \ref{Bal_angle} shows the data pointing away from the town and the lower row (c, d) facing town. The blue dashed line in the plots show the luminance before and the black solid line after the lights were switched off. The data shows that the relative change in luminance is highest near zenith.

\begin{figure}[h]
\centering
\includegraphics[width=0.8\columnwidth]{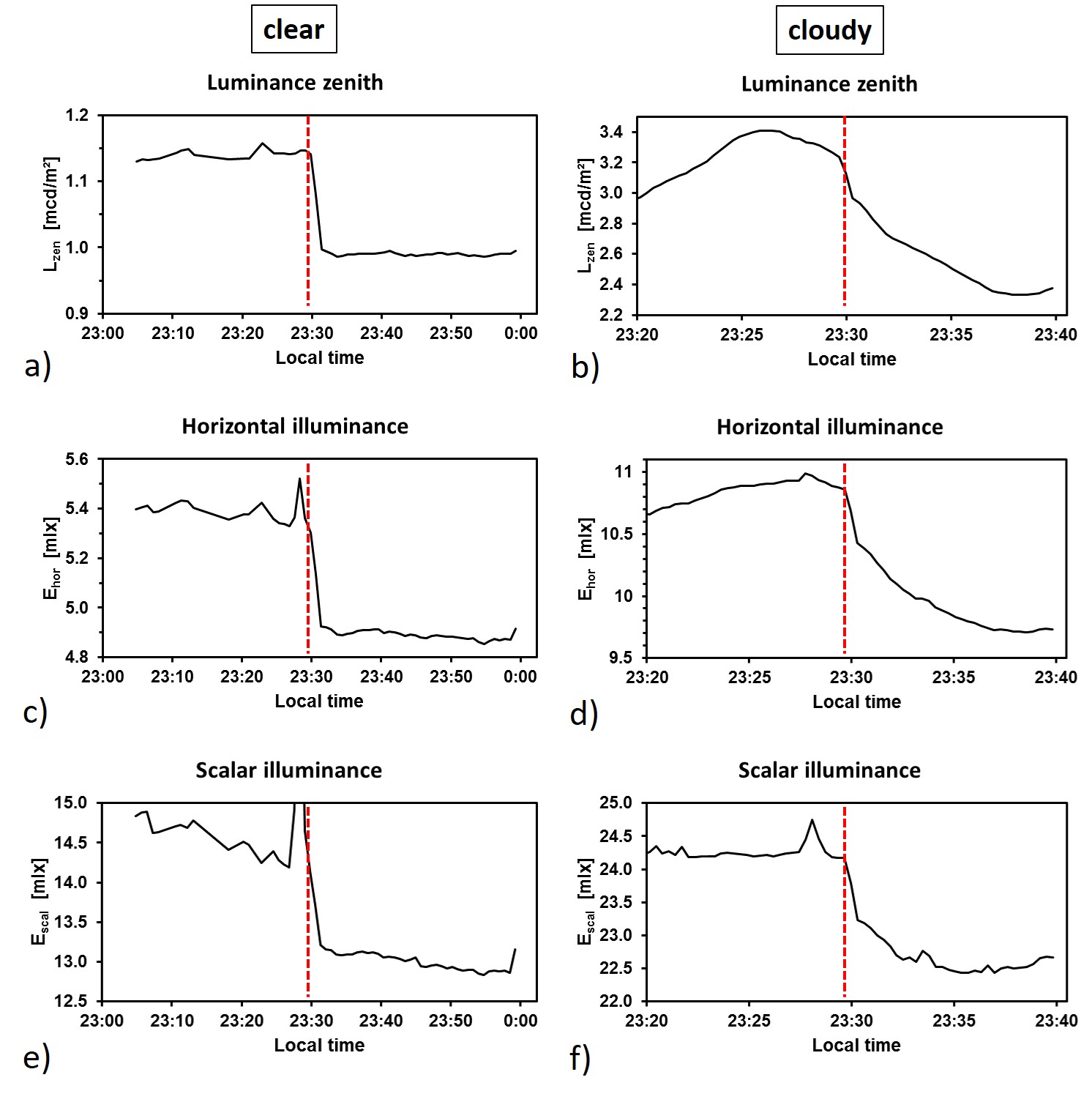}
\caption{Time series obtained in Balaguer showing luminance at zenith (a, b), horizontal illuminance (c, d) and scalar illuminance (e, f) extracted from the images. The left column shows data for clear and right column for cloudy sky. The red dashed lines in each plot indicate the switch-off scheduled at 23:30 for both nights.}
\label{Bal_time}
\end{figure}

\subsubsection*{Temporal change in luminance and illuminance}
During both nights, the switch-off was observed as part of a longer time series. During the clear night, data was obtained from 23:06 to 00:32 local time and during the cloudy night images were taken from 23:08 to 23:39 local time. ISO was set to 1600 for both nights and exposure time was 30s for clear and 15s for cloudy night.

Figure \ref{Bal_time} shows the temporal change in luminance at zenith (a, b), horizontal illuminance (c, d) and scalar illuminance (e, f) with the left column for clear and right column for cloudy sky. The switch-off scheduled at 23:30 for both nights (indicated by the red dashed lines in each plot) is apparent in all extracted values for the clear night, showing a sharp drop-off at the scheduled time. For the clear night, the zenith luminance (Fig. \ref{Bal_time} a) is relatively stable with some fluctuations mainly due to changed atmospheric conditions (there were a few small clouds at the beginning of data taking). Overall, a slight positive slope indicates an increase in luminance. Both the horizontal illuminance (Fig. \ref{Bal_time} c) and scalar illuminance (Fig. \ref{Bal_time} e) rather have negative slopes during the clear night, more pronounced for the scalar illuminance \footnote{The sharp peak just before the switch-off in clear night illuminance was caused by a car passing by.}. This is consistent with previously published observations as lights tend to be switched off as the night progresses \cite{Kyba:2015_isqm}.

For the data set obtained during cloudy conditions shown in Fig. \ref{Bal_time} (b, d, f) the change caused by the lamp switch-off is not as pronounced as in the clear night data set in all extracted values. Here, the fluctuation of the NSB and illuminance due to changing clouds cover was relatively high. These fluctuations were highest for zenith luminance and horizontal illuminance because the changing clouds are mostly affecting the values near zenith, while the scalar illuminance remained more stable. Nevertheless, the drop-off in brightness can still be seen in all three data sets and is best visible in the scalar illuminance Fig. \ref{Bal_time} (f).

\subsection*{3.2. Public lights in \`{A}ger (organized switch-off)}
\subsubsection*{Luminance maps}
\begin{figure}[tp]
\centering
\includegraphics[width=0.75\columnwidth]{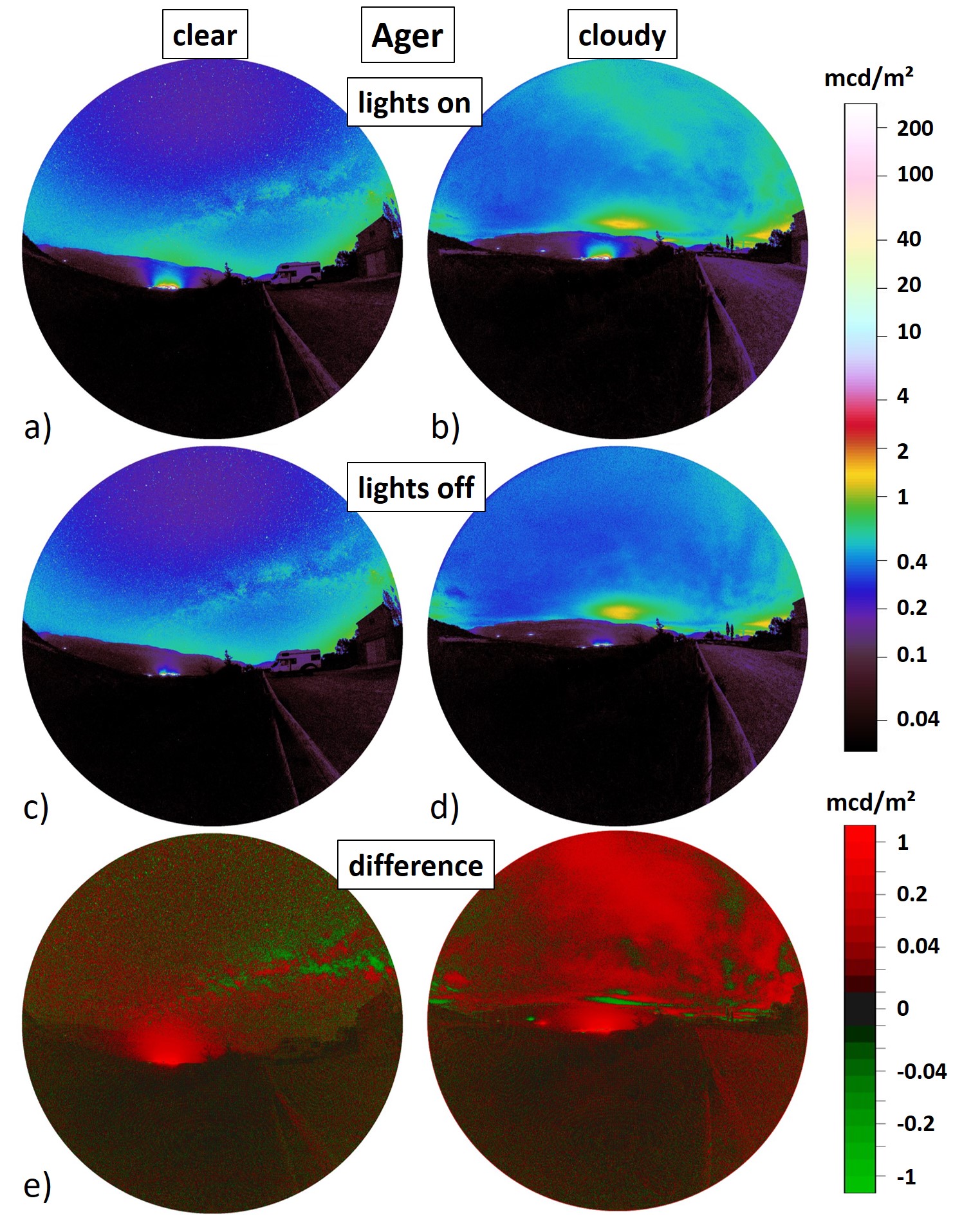}
\caption{Excerpt from the data obtained for the organized switch-off of public lights in \`{A}ger, Spain. Left column (a, c, e) shows clear night data and right hand column (b, d, f) shows cloudy night data. The upper row (a, b) shows vertical luminance maps before the switch-off, middle row (c, d) vertical luminance maps after the switch-off and lower row shows differential luminance maps. See text for more details on times and location.}
\label{Ager_sub}
\end{figure}

Figure \ref{Ager_sub} shows part of the data obtained from Port d' \`{A}ger observing the manual switch-off in the village of \`{A}ger. Here the camera was pointed approximately towards the horizon. The left column (a, c, e) shows data from the clear night and right column (b, d, f) data from the cloudy night. The upper row (a, b) shows luminance maps before the public lights were switched off with data obtained on May 4th 2016 starting at 02:39:15 and on May 6th 2016 at 01:59:32 local time, respectively. The middle row (c, d) shows luminance maps after the light has been switched off, obtained at May 4th 2016, 02:52:27 and on May 6th 2016 at 02:16:34 local time, respectively. The lower row shows differential luminance maps, obtained by subtracting the images after the switch-off (middle rows) from the images before the switch-off (upper rows).

In the vertical luminance maps (Fig. \ref{Ager_sub} a, c) obtained in the night with clear sky, the light from the village of \`{A}ger is the dominant light source. Skyglow is concentrated to the horizon and the Milky Way is clearly visible. No obvious change in the sky brightness is apparent from the clear night luminance maps, while the switch-off of the public light is. This impression is supported by the differential image shown in the lower row Fig. \ref{Ager_sub} (e), where red indicates darkening and green indicates brightening. For the clear night data (Fig. \ref{Ager_sub} e), the sky luminance and the ground luminance have not significantly changed, as the changes are mainly due to the rotation of the Earth and apparent movement of stars in the images resulting in about equally sized green and red regions. The switch-off of the lights from \`{A}ger can be seen by the red region near the village, as expected.

In the cloudy sky luminance maps (Fig. \ref{Ager_sub} b, d), the lights from \`{A}ger can be spotted and a bright skyglow is clearly visible above the mountain in the center of the images. This skyglow most likely originates from the town of Tremps (ca. 6000 inhabitants, distance ca. 25 km, zenith NSB ca. 0.59 mcd/m$^2$ according to \cite{falchi2016supplement}). For a map see Fig. \ref{map} right hand side and Fig. \ref{WA} for the NSB data from Falchi et al. \cite{falchi2016WA} in Methods section. Furthermore, a change in NSB near zenith is also apparent when comparing the two luminance maps (see upper center of Fig. \ref{Ager_sub} b and d). The differential image for the cloudy night in Fig. \ref{Ager_sub} (f) clearly shows a bigger change in brightness than for the clear night. The sky luminance has clearly changed both due to moving clouds and switching off of the lights. The majority of red regions indicate an overall darkening with lights being switched off, with only few green patches. Furthermore, a change in ground illuminance can be seen in the lower right corner of the subtracted image.

\begin{figure}[htbp!]
\centering
\includegraphics[width=0.35\columnwidth]{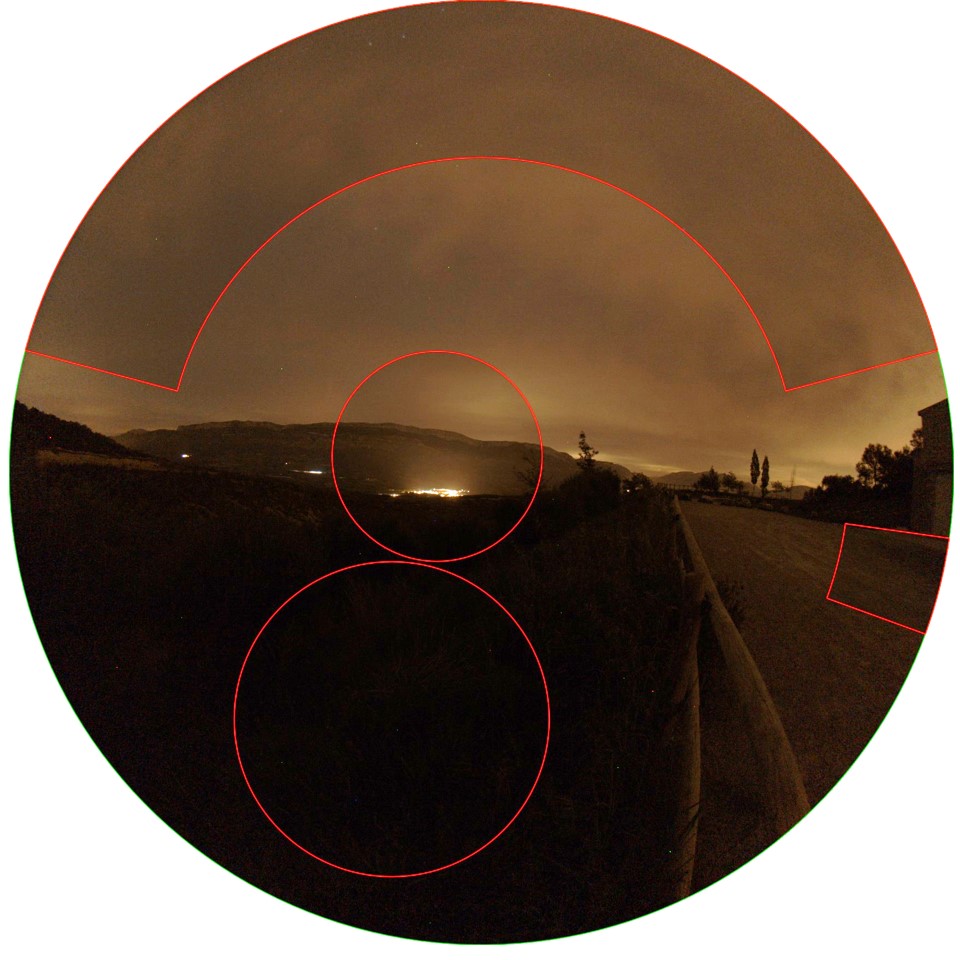}
\caption{RGB image of \`{A}ger obtained during the cloudy night after the switch-off. The red regions indicate the analyzed luminance data.}
\label{Ager_RGB}
\centering
\includegraphics[width=0.8\columnwidth]{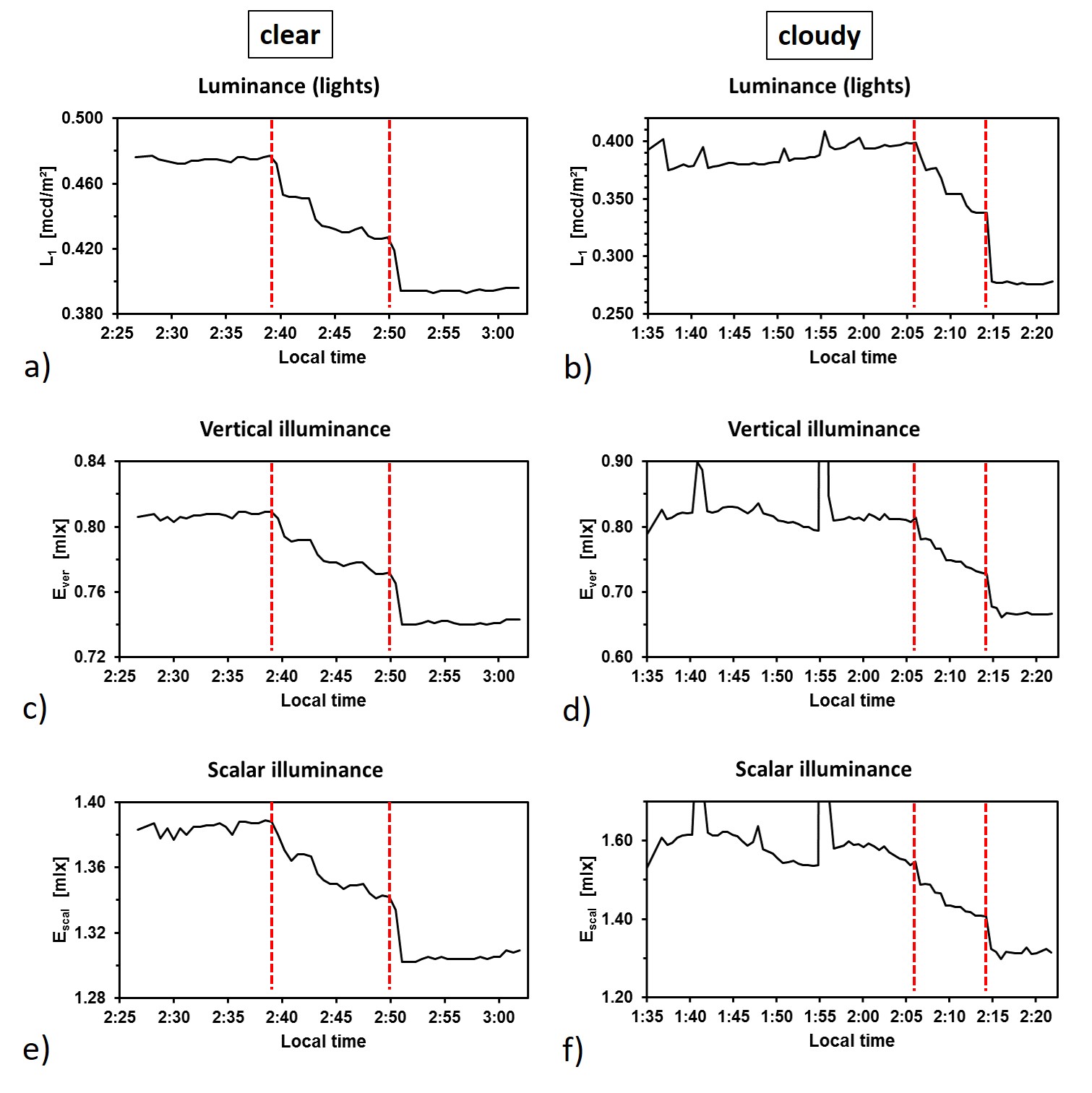}
\caption{Temporal evolution of the luminance near the lights of \`{A}ger (a, b), the change in vertical illuminance (c, d) and in scalar illuminance (e, f). Left hand side shows clear night and right hand side cloudy night data. Red dashed lines indicate start and end of the four step switch-off sequence.}
\label{Ager_time1}
\end{figure}

\begin{figure}[h]
\centering
\includegraphics[width=0.8\columnwidth]{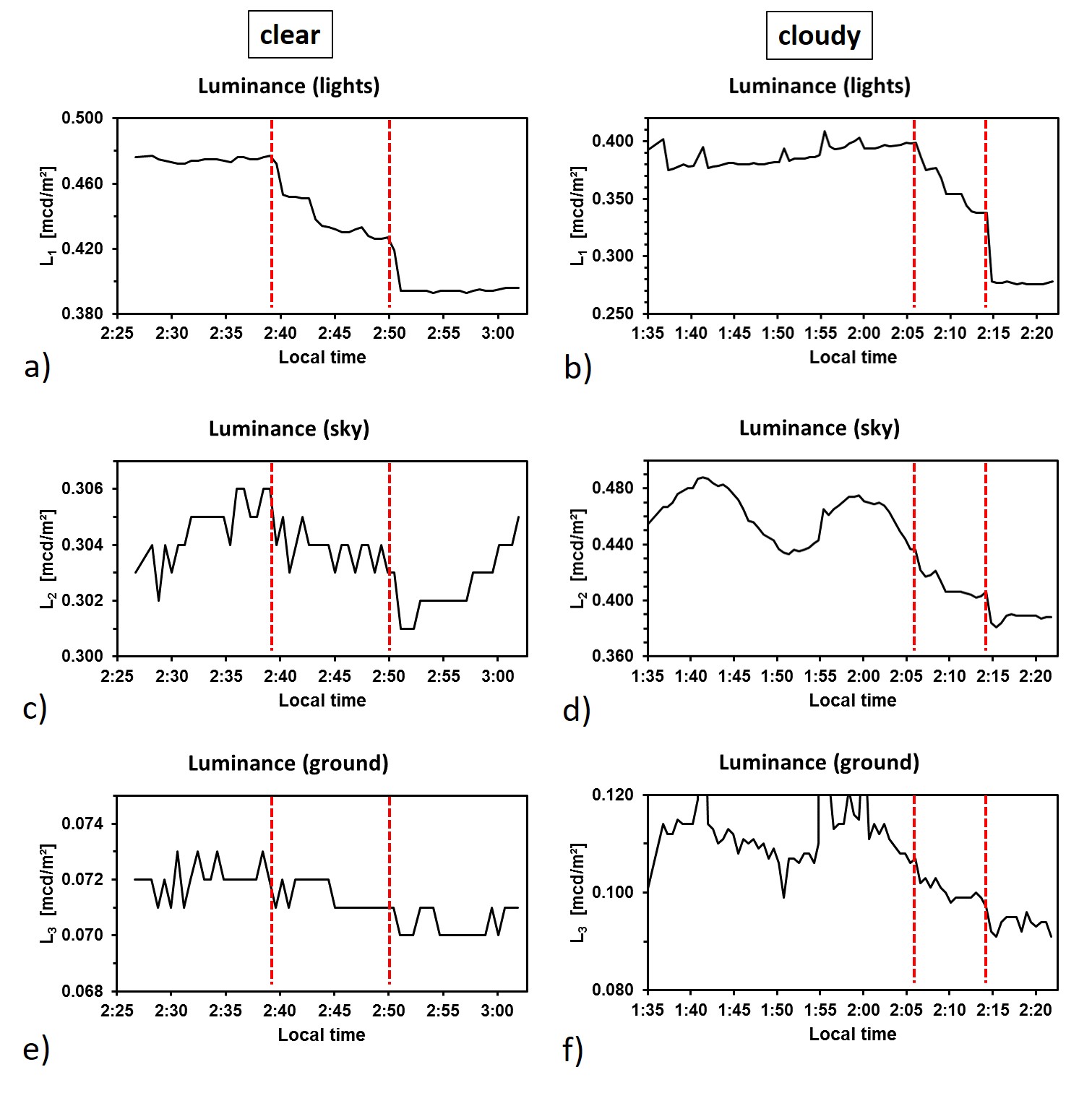}
\caption{Temporal evolution of the luminance near the lights of \`{A}ger (a, b), the change in sky luminance (c, d) and in ground luminance (e, f). Left hand side shows clear night and right hand side cloudy night data. Red dashed lines indicate start and end of the four step switch-off sequence.}
\label{Ager_time2}
\end{figure}

\pagebreak
\subsubsection*{Temporal change in luminance and illuminance}
The switch-off of public lights in \`{A}ger was performed manually at request in four sequences identical for both nights. The start and end of the sequence is indicated by the red dashed lines in each following plot (Fig. \ref{Ager_time1} and Fig. \ref{Ager_time2}). During both nights, the switch-off was observed as part of a time series of images. During the clear night, data was obtained from 02:27 to 03:10 local time, and during the cloudy night from 01:35 to 02:21 local time. ISO was set to 3200 and exposure time was 30s for both nights.

The luminance was analyzed for different regions of interest. Figure \ref{Ager_RGB} shows the defined regions. The luminance near \`{A}ger was defined by a 20$^{\circ}$ circle around the main lights. The luminance of the sky was defined by an area in the upper image spanning from 60$^{\circ}$ to 90$^{\circ}$ polar angle and 285$^{\circ}$ to 75$^{\circ}$ azimuth angle. And the luminance on the ground (gravel road pavement) was defined as an area in the lower right part of the image spanning from 75$^{\circ}$ to 90 $^{\circ}$ polar angle and 255$^{\circ}$ to 265$^{\circ}$ azimuth angle. The luminance on the grass by a 20$^{\circ}$ circle. 

Figure \ref{Ager_time1} shows the temporal evolution of the luminance near the lights of \`{A}ger (a, b), the change in vertical illuminance (c, d) and in scalar illuminance (e, f). Left hand side shows clear night and right hand side cloudy night data. The sequence of the four switch-offs is apparent in the luminance near the village (lights - Fig. \ref{Ager_time1} a, b) for both nights. For the clear night, the sequence is also apparent in both illuminance data sets (Fig. \ref{Ager_time1} c, e). For the cloudy night, the fluctuations in the illuminance values make it harder to spot all four steps cleary. Nevertheless, start and end of the sequence are apparent in both illuminance data sets (Fig. \ref{Ager_time1} d, f).

Figure \ref{Ager_time2} shows again the temporal evolution of the luminance near the lights of \`{A}ger (a, b), followed by the luminance of the sky (c, d) and luminance on the ground (e, f). Again, clear sky data is shown in the left column and cloudy sky data in the right column.

For the clear night, the sequence is not apparent in the sky or ground luminance data (Fig. \ref{Ager_time2} c, e). The changes due to the milky way moving slowly into the image and other fluctuations mask the pattern. By close inspection, the sky luminance (Fig. \ref{Ager_time2} c) might show a weak signal at the start and end of the sequence. However, the signal changes by only 5 $\mu$cd/m$^2$ throughout the measurement interval (The quantization of the camera is $\approx$ 1 $\mu$cd/m$^2$ at the chosen settings). No such signal is detectable in the ground luminance (Fig. \ref{Ager_time2} e), when the signal changes only by 3 $\mu$cd/m$^2$ over the measurement interval. For the cloudy night, the first, second and last step of the four step switch-off sequence is apparent in sky luminance and ground luminance (Fig. \ref{Ager_time2} d, f), although the individual values fluctuate a lot due to changing cloud cover.
\begin{figure}[tp]
\centering
\includegraphics[width=0.4\columnwidth]{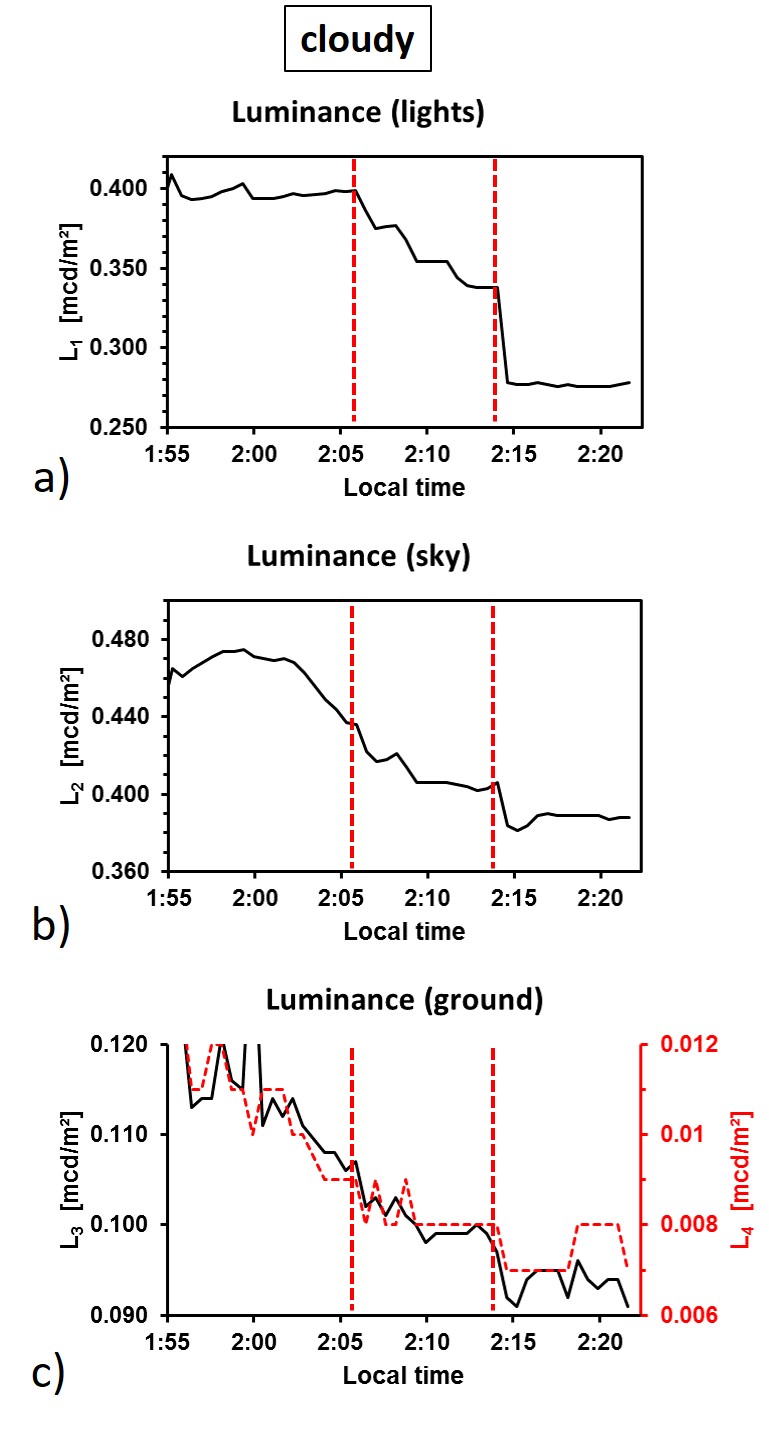}
\caption{Temporal evolution of the NSB for the cloudy night on a shorter time scale. (a) Luminance near the lights of \`{A}ger, (b) sky luminance and (c) luminance of the ground (black line - pavement, red line - grass floor).}
\label{Ager_time3}
\end{figure}

The surfaces near the vantage point are more strongly illuminated on the cloudy than on the clear night. Therefore, we also analyzed the ground in the lower center of the image (grass covered) that had a lower albedo than the pavement but only for the cloudy night. Figure \ref{Ager_time3} shows this data with the luminance near the lights of \`{A}ger as reference (a), the sky luminance (b) and the ground luminance from the pavement (c, black solid line) and the luminance of the grass floor (c, red dashed line). The data is shown for a shorter time frame than the full data set. The pattern for the luminance reflected from the ground is similar for both investigated regions (grass floor and pavement). Very low luminance values on the order of a few $\mu$cd/m$^2$ were measured of light reflected from the grass floor. It has to be noted, that these values are almost near the noise floor of the camera (quantization is $\approx$ 1 $\mu$cd/m$^2$) for the chosen exposure time and the accuracy of the camera in this region is not known. Nevertheless, differential photometry with simultaneous tracking of a reference signal (here the lights of \`{A}ger) seems to be capable to detect such low values. A source of error could be stray light or lens flare effects originating e.g. from imperfections of the anti-reflection coating, scratches or dust etc. Nevertheless, a precise calibration at such low values appears to be difficult but not impossible.

\section*{4. Summary and conclusion}
Differential photometry is an excellent tool to observe the dynamics of ALAN and therefore also skyglow. It makes it possible to unravel small changes in night sky brightness and illuminance levels simultaneously, by taking time series of images. When simultaneously tracking a reference signal, very small changes in luminance or illuminance can be detected.

We used a commercial DSLR camera with fisheye lens to observe an automated regular switch-off of ornamental light in the town of Balaguer and an organized switch-off of public lights in the village of \`{A}ger, both near Montsec Astronomical Park in Spain. Both sites were investigated during two nights with clear and cloudy conditions. The images were analyzed with the Sky Quality Camera software (Euromix, Ljubljana, Slovenia). Luminance maps, subtracted images, illuminance and luminance values for specific regions were investigated.

By taking all-sky images, we find that during the clear night the ornamental lights in Balaguer have nearly 21$\%$ contribution to the artificial skyglow at zenith at the observational site, while the impact on illuminance values was lower. Furthermore, the influence of the switch-off on the observed values was lower during the cloudy night compared to the clear night.

We are able to track very small changes in night sky brightness and ground luminance during a cloudy night near \`{A}ger by taking vertical images. Changes in luminance values on the order of a few $\mu$cd/m$^2$ have been observed, which is below the uncertainty range of the initial calibration. By simultaneously analyzing a region of interest and the lights directly, we are able to link the small changes to the switch-offs. This was not possible for the clear night.

In conclusion, differential photometry with wide-angle lenses promises to be a good tool for analyzing the contribution of individual lights to the night sky brightness or the ground illuminance. Bara et al. have done a very interesting analysis linking SQM measurements with camera time series \cite{bara2017estimating}. A comparable measurement could be feasible with one camera alone. A drawback of using a DSLR camera with fisheye at low light levels is the rather long integration time, but this is more than made up for by the advantage of the simultaneous measurement of many parameters, greatly easing the interpretation. 

It has to noted, that the time series were part of a transect from Balaguer to Port d' \`{A}ger \cite{jechow2017balaguer}, which was the main aim of the data acquisition. Further studies should include long term observations for example of a full night. It would be interesting to test the boundaries of fisheye lens observations and high frequency long-term measurements for low light levels. Further work on calibration at low light levels and tests of the long term stability of the method and impact of environmental parameters (e.g. temperature, moisture) will be necessary to increase the accuracy.

\section*{Acknowledgements}
We thank the municipalities of \`{A}ger and Balaguer for their cooperation permitting and assisting with public and ornamental light switch-offs. We are also thankful to all the participants at the Stars4All/LoNNe intercomparison that supported this work, specifically Guillem Marti and Pol Massana for driving the car on the first day of the transect, and Constantinos Bouroussis and Henk Spoelstra for joining the transect on the second night. We thank the ceilometer research team of IDAEA - CSIC (Instituto de Diagnóstico Ambiental y Estudios del Agua, Spanish Council for Scientific Research) for providing processed data of the instrument at PAM-COU.

\section*{Funding}
Andreas Jechow is supported by the ILES project funded by the Leibniz Association, Germany (SAW-2015-IGB-1). We thank the STARS4ALL awareness platform that funded the intercomparison campaign. STARS4ALL is a project funded by the European Union H2020-ICT-2015-688135. This article is based upon work from COST Action ES1204 LoNNe, supported by COST (European Cooperation in Science and Technology).
\pagebreak
\appendix
\section{SQM night sky brightness measurements for 2016}
\begin{figure}[htbp]
\centering
\includegraphics[width=0.6\columnwidth]{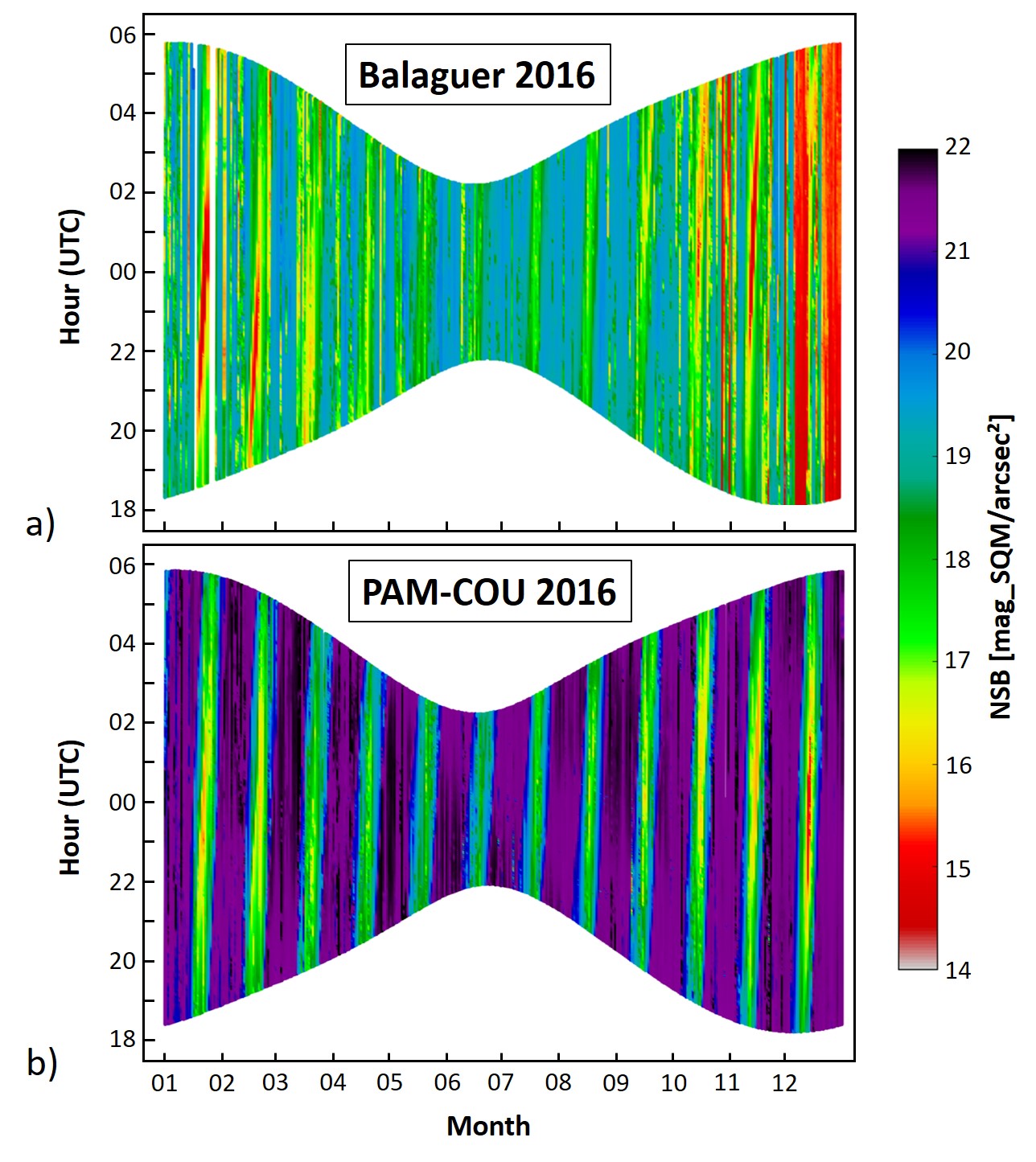}
\caption{Annual evolution of the NSB  for the year 2016 measured with SQMs for a) in Balaguer and b) at PAM-COU near \`{A}ger .}
\label{app1}
\end{figure}

Figure \ref{app1} shows the evolution of the NSB for the year 2016 measured with SQMs \footnote{The NSB measured with an SQM can be approximately converted to luminance by using the equation: $L_v \approx 10.8 \cdot 10^4 \cdot 10^{-0.4mag_{SQM}}$. Please see \cite{hanel2017measuring} for a brief discussion.} a) in Balaguer and b) at PAM-COU near \`{A}ger. Data is taken every 45 s. In both plots lunar cycles are clearly visible. The NSB brightness at PAM-COU for moonless nights is between 21 and 22 mag$_{SQM}$/arcsec$^2$ with cloudy periods sometimes reaching very low values. The NSB brightness in Balaguer for moonless nights is between 19 and 20 mag$_{SQM}$/arcsec$^2$ with cloudy periods sometimes reaching very bright values of 15 mag$_{SQM}$/arcsec$^2$.

\section{SQM night sky brightness measurements for the observation nights}
\begin{figure}[htbp]
\centering
\includegraphics[width=0.9\columnwidth]{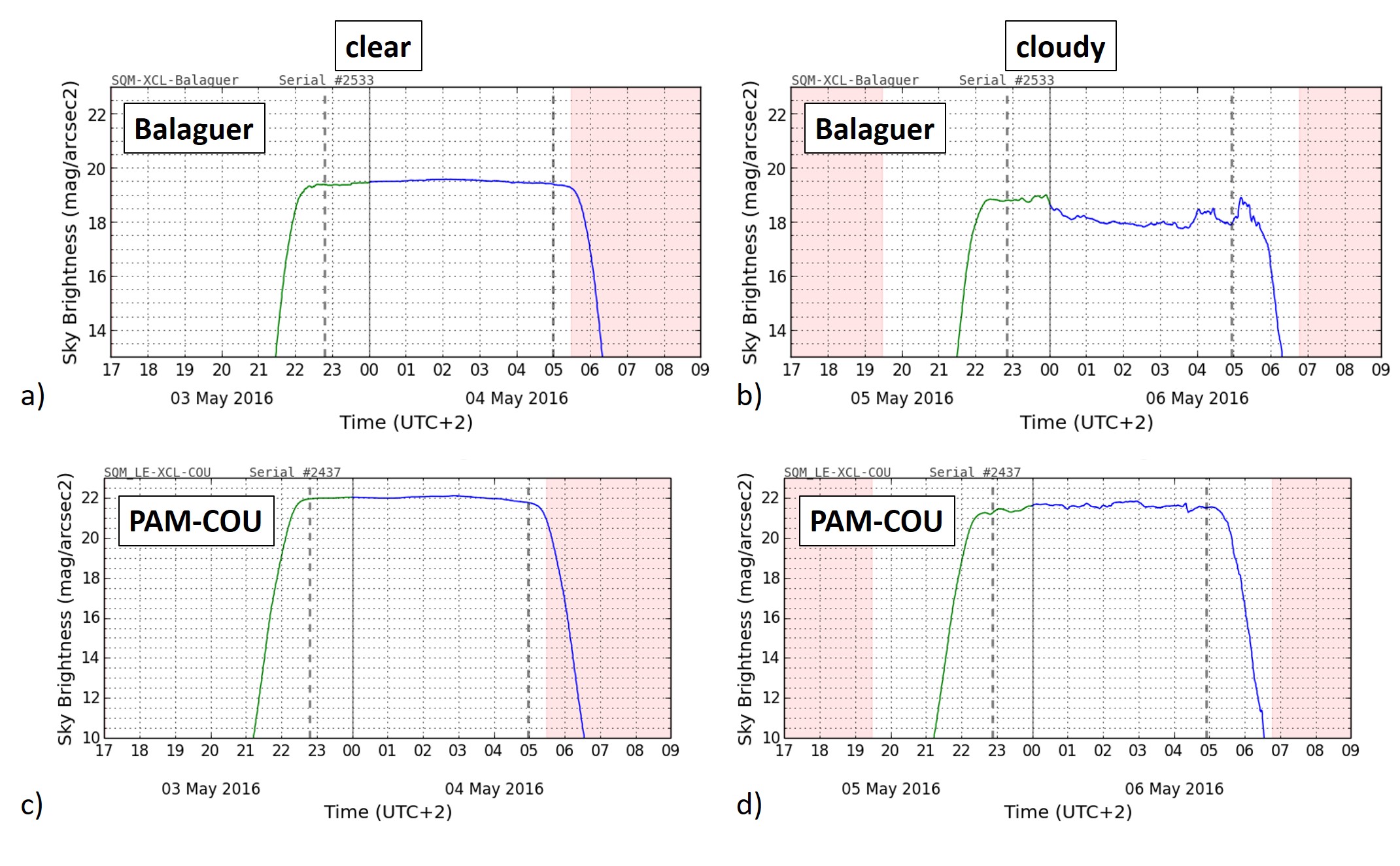}
\caption{Evolution of the NSB for the two nights of photometry data taking measured with SQMs in Balaguer (upper row) and at PAM-COU near \`{A}ger (lower row).}
\label{app2}
\end{figure}

Figure \ref{app2} shows the evolution of the NSB measured with SQMs for the nights of data taking with the DSLR camera. The left column (a, c) shows clear night data, right column (b, d) cloudy night data. Upper row (a, b) shows data taken in Balaguer  and lower row (c, d) data taken at PAM-COU near \`{A}ger.

%% References

\pagebreak

\end{document}